\documentclass[12pt,reqno]{amsart}%
\usepackage{graphicx}
\usepackage{amscd}
\usepackage{amsmath}
\usepackage{epsfig}
\usepackage{amsfonts}
\usepackage{amssymb}%
\providecommand{\U}[1]{\protect\rule{.1in}{.1in}}
\numberwithin{equation}{section}
\providecommand{\U}[1]{\protect\rule{.1in}{.1in}}
\providecommand{\U}[1]{\protect\rule{.1in}{.1in}}
\allowdisplaybreaks

\theoremstyle{plain}

\begin{document}
\title[Photon Helicity]{The Pauli-Luba\'{n}ski Vector, Complex Electrodynamics,\\and Photon Helicity}
\author{Sergey I. Kryuchkov }
\address{School of Mathematical and Statistical Sciences, Arizona State University,
Tempe, AZ 85287--1804, U.S.A.}
\email{sergeykryuchkov@yahoo.com}
\author{Nathan A. Lanfear}
\address{School of Mathematical and Statistical Sciences, Arizona State University,
Tempe, AZ 85287--1804, U.S.A.}
\email{nlanfear@asu.edu}
\author{Sergei~K.~Suslov}
\address{School of Mathematical and Statistical Sciences \& Mathematical, Computational
and Modeling Sciences Center, Arizona State University, Tempe, AZ 85287--1804, U.S.A.}
\email{sks@asu.edu}
\urladdr{http://hahn.la.asu.edu/\symbol{126}suslov/index.html}
\dedicatory{Dedicated to Margarita A.~Man'ko and Vladimir I.~Man'ko on the occasion of
their $75+75=150$ birthday\\for their great contributions to science and pedagogy.}\date{\today}

\begin{abstract}
We critically analyze the concept of photon helicity and its connection with
the Pauli-Luba\'{n}ski vector from the viewpoint of the complex
electromagnetic field, $\mathbf{E}+i\mathbf{H},$ sometimes attributed to
Riemann but studied by Weber, Silberstein and Minkowski. To this end, a
complex covariant form of Maxwell's equations is used.

\end{abstract}
\maketitle

\section{Introduction}

All physically interesting unitary ray representations of the proper
orthochronous inhomogeneous Lorentz group (known nowadays as the Poincar\'{e}
group) were classified by Wigner \cite{Wigner39} and, since then, this
approach has been utilized for the mathematical description of mass and spin
of an elementary particle. By definition, the Pauli-Luba\'{n}ski pseudo-vector
is given by
\begin{equation}
w_{\mu}=\frac{1}{2}e_{\mu\nu\sigma\tau}p^{\nu}M^{\sigma\tau},\qquad p_{\mu
}w^{\mu}=0, \label{PauliLubanskiVector}%
\end{equation}
where $p_{\mu}$ is the relativistic linear momentum operator and
$M^{\sigma\tau}$ are the corresponding angular momentum operators. The mass
and spin of a particle are defined in terms of two quadratic invariants
(Casimir operators of the Poincar\'{e} group) as follows%
\begin{equation}
p^{2}=p_{\mu}p^{\mu}=m^{2},\qquad w^{2}=w_{\mu}w^{\mu}=-m^{2}s\left(
s+1\right)  ,\qquad m>0 \label{MassSpin}%
\end{equation}
(see, for example, \cite{Bargmann54}, \cite{BargmannWigner48},
\cite{Bogolubovetal90}, \cite{Kuzetal}, \cite{Lub41}, \cite{Lub42-I},
\cite{Lub42-II}, \cite{LomontMoses62}, \cite{RumerFetQFT}, \cite{Ryder},
\cite{ScheckQuantumPhysics} and the references therein; we use the standard
notations that are introduced in the body of the article).

For the massless fields, when $m=0,$ one gets $w^{2}=p^{2}=pw=0,$ and the
Pauli-Luba\'{n}ski vector should be proportional to $p:$%
\begin{equation}
w_{\mu}=\lambda p_{\mu} \label{WrongHelicity}%
\end{equation}
(acting on the corresponding eigenstates \cite{Naber12}, \cite{Ryder}). The
number $\lambda$ is called the helicity of the representation and the value
$s=\left\vert \lambda\right\vert $ is sometimes called the spin of a particle
with zero mass \cite{Bogolubovetal90}, \cite{Ryder},
\cite{ScheckQuantumPhysics}. One of the goals of this article is to show that,
in the case of the electromagnetic field, the constant $\lambda$ in the latter
equation is fixed, otherwise violating the classical Maxwell equations.

As a result, instead of being given by the constant of proportionality in
relation (\ref{WrongHelicity}), the helicity of the photon should be defined,
as it is traditionally done in particle physics, by $\lambda=\mathbf{k}%
\cdot\mathbf{M}/k_{0},$ where $k=\left(  k_{0},\mathbf{k}\right)  $ and
$\mathbf{M}$ is the photon angular momentum ($k^{2}=k_{0}^{2}-\mathbf{k}^{2}$
$=0).$ But one needs a proper realization of the action of these operators on
the photon field tensor in covariant form \cite{MinkowskiI},
\cite{MinkowskiII}; or, in $3D$-form, on the complex electromagnetic field
vector $\mathbf{F}=\mathbf{E}+i\mathbf{H}$ discussed in \cite{Barut80},
\cite{Bat15}, \cite{Bia96}, \cite{Bia:Bia13}, \cite{LanLif2}, \cite{LapUh31},
\cite{MinkowskiI}, \cite{RumerFetQFT}, \cite{Schwinger}, \cite{SilbI},
\cite{SilbII}, \cite{Taylor52}, \cite{Weber}. The sign of the constant
$\lambda=$ $\pm1$ is fixed then by a \textquotedblleft
continuity\textquotedblright, in view of the invariance of the upper(lower)
light cone under a proper Lorentz transformation (see, for example,
\cite{Bogolubovetal90}, \cite{Kuzetal}, and \cite{TroTy94}).

\section{Maxwell's Equations in Complex and Covariant Forms}

We would like to introduce two complex vector fields%
\begin{equation}
\mathbf{F}=\mathbf{E}+i\mathbf{H},\qquad\mathbf{G}=\mathbf{D}+i\mathbf{B}
\label{FGcomplex}%
\end{equation}
($\mathbf{E}$ is the electric field, $\mathbf{D}$ the electric displacement
field, $\mathbf{H}$ the magnetic field, and $\mathbf{B}$ the magnetic
induction field) and present the phenomenological Maxwell equations in a
compact form,%
\begin{equation}
\frac{i}{c}\left(  \frac{\partial\mathbf{G}}{\partial t}+4\pi\mathbf{j}%
\right)  =\operatorname{curl}\mathbf{F},\qquad\mathbf{j}=\mathbf{j}^{\ast},
\label{MaxwellOne}%
\end{equation}%
\begin{equation}
\operatorname{div}\mathbf{G}=4\pi\rho,\qquad\rho=\rho^{\ast},
\label{MaxwellTwo}%
\end{equation}
where the asterisk stands for complex conjugation. (We are working in Gaussian
units; complex-valued $\rho$ and $\mathbf{j}$ may be related to the free
magnetic charge and current, which have not been observed yet in nature
\cite{Bia:Bia75}, \cite{Schwinger}.)

With the help of the complex fields $\mathbf{F}=\mathbf{E}+i\mathbf{H}$ and
$\mathbf{G}=\mathbf{D}+i\mathbf{B},$ we introduce the following anti-symmetric
4-tensor,%
\begin{equation}
Q^{\mu\nu}=-Q^{\nu\mu}=\left(
\begin{array}
[c]{cccc}%
0 & -G_{1} & -G_{2} & -G_{3}\\
G_{1} & 0 & iF_{3} & -iF_{2}\\
G_{2} & -iF_{3} & 0 & iF_{1}\\
G_{3} & iF_{2} & -iF_{1} & 0
\end{array}
\right)  \label{ComplexFieldTensor}%
\end{equation}
and use the standard 4-vector notation, $x^{\mu}=\left(  ct,\mathbf{r}\right)
$ and $j^{\mu}=\left(  c\rho,\mathbf{j}\right)  $ for contravariant
coordinates and current, respectively.

Maxwell's equations then take the covariant form \cite{KrLanSu15},
\cite{LapUh31}:%
\begin{equation}
\frac{\partial}{\partial x^{\nu}}Q^{\mu\nu}=-\frac{\partial}{\partial x^{\nu}%
}Q^{\nu\mu}=-\frac{4\pi}{c}j^{\mu}\label{MaxwellCovariant}%
\end{equation}
with summation over two repeated indices. Indeed, in a block form,%
\begin{equation}
\frac{\partial Q^{\mu\nu}}{\partial x^{\nu}}=\frac{\partial}{\partial x^{\nu}%
}\left(
\begin{array}
[c]{cc}%
0 & -G_{q}\medskip\\
G_{p} & ie_{pqr}F_{r}%
\end{array}
\right)  =\left(
\begin{array}
[c]{c}%
-\operatorname{div}\mathbf{G}=-4\pi\rho\medskip\\
\dfrac{1}{c}\dfrac{\partial\mathbf{G}}{\partial t}+i\operatorname{curl}%
\mathbf{F}=-\dfrac{4\pi}{c}\mathbf{j}%
\end{array}
\right)  ,\label{MaxwellCavariantExplicit}%
\end{equation}
which verifies this fact.

The continuity equation,%
\begin{equation}
0\equiv\frac{\partial^{2}Q^{\mu\nu}}{\partial x^{\mu}\partial x^{\nu}}%
=-\frac{4\pi}{c}\frac{\partial j^{\mu}}{\partial x^{\mu}},
\label{CovariantContinuityEquation}%
\end{equation}
or, in the $3D$-form,%
\begin{equation}
\frac{\partial\rho}{\partial t}+\operatorname{div}\mathbf{j}=0, \label{ContEq}%
\end{equation}
describes conservation of the electrical charge.

\section{Dual Complex Field Tensors}

Two dual anti-symmetric field tensors of the complex fields, $\mathbf{F}%
=\mathbf{E}+i\mathbf{H}$ and $\mathbf{G}=\mathbf{D}+i\mathbf{B},$ are given
by
\begin{align}
Q^{\mu\nu}  &  =\left(
\begin{array}
[c]{cccc}%
0 & -G_{1} & -G_{2} & -G_{3}\\
G_{1} & 0 & iF_{3} & -iF_{2}\\
G_{2} & -iF_{3} & 0 & iF_{1}\\
G_{3} & iF_{2} & -iF_{1} & 0
\end{array}
\right) \label{ComplexQ}\\
&  =\left(
\begin{array}
[c]{cccc}%
0 & -D_{1} & -D_{2} & -D_{3}\\
D_{1} & 0 & -H_{3} & H_{2}\\
D_{2} & H_{3} & 0 & -H_{1}\\
D_{3} & -H_{2} & H_{1} & 0
\end{array}
\right)  +i\left(
\begin{array}
[c]{cccc}%
0 & -B_{1} & -B_{2} & -B_{3}\\
B_{1} & 0 & E_{3} & -E_{2}\\
B_{2} & -E_{3} & 0 & E_{1}\\
B_{3} & E_{2} & -E_{1} & 0
\end{array}
\right) \nonumber
\end{align}
and%
\begin{align}
P_{\mu\nu}  &  =\left(
\begin{array}
[c]{cccc}%
0 & F_{1} & F_{2} & F_{3}\\
-F_{1} & 0 & iG_{3} & -iG_{2}\\
-F_{2} & -iG_{3} & 0 & iG_{1}\\
-F_{3} & iG_{2} & -iG_{1} & 0
\end{array}
\right) \label{ComplexP}\\
&  =\left(
\begin{array}
[c]{cccc}%
0 & E_{1} & E_{2} & E_{3}\\
-E_{1} & 0 & -B_{3} & B_{2}\\
-E_{2} & B_{3} & 0 & -B_{1}\\
-E_{3} & -B_{2} & B_{1} & 0
\end{array}
\right)  +i\left(
\begin{array}
[c]{cccc}%
0 & H_{1} & H_{2} & H_{3}\\
-H_{1} & 0 & D_{3} & -D_{2}\\
-H_{2} & -D_{3} & 0 & D_{1}\\
-H_{3} & D_{2} & -D_{1} & 0
\end{array}
\right)  .\nonumber
\end{align}
The real part of the latter represents the standard electromagnetic field
tensor in a medium \cite{Barut80}, \cite{Pauli}, \cite{ToptyginII}. As for the
imaginary part of (\ref{ComplexQ}), which, ironically, Pauli called an
\textquotedblleft artificiality\textquotedblright\ in view of its non-standard
behavior under spatial inversion \cite{Pauli}, the use of complex conjugation
restores this important symmetry for our complex field tensors.

In the complex case under consideration, the dual tensor identities are given
by%
\begin{equation}
e_{\mu\nu\sigma\tau}Q^{\sigma\tau}=2iP_{\mu\nu},\qquad2iQ^{\mu\nu}=e^{\mu
\nu\sigma\tau}P_{\sigma\tau}. \label{DualPQ}%
\end{equation}
Here, $e^{\mu\nu\sigma\tau}=-e_{\mu\nu\sigma\tau}$ and $e_{0123}=+1$ is the
Levi-Civita symbol \cite{Fock64}. Then%
\begin{equation}
6i\frac{\partial Q^{\mu\nu}}{\partial x^{\nu}}=e^{\mu\nu\lambda\sigma}\left(
\frac{\partial P_{\lambda\sigma}}{\partial x^{\nu}}+\frac{\partial
P_{\nu\lambda}}{\partial x^{\sigma}}+\frac{\partial P_{\sigma\nu}}{\partial
x^{\lambda}}\right)  \label{DualDiffIdentity}%
\end{equation}
and both pairs of Maxwell's equations can also be presented in the form
\cite{KrLanSu15}:%
\begin{equation}
\frac{\partial P_{\mu\nu}}{\partial x^{\lambda}}+\frac{\partial P_{\nu\lambda
}}{\partial x^{\mu}}+\frac{\partial P_{\lambda\mu}}{\partial x^{\nu}}%
=-\frac{4\pi i}{c}e_{\mu\nu\lambda\sigma}j^{\sigma}
\label{CovariantMaxwellTwo}%
\end{equation}
in addition to the one given above,%
\begin{equation}
\frac{\partial Q^{\mu\nu}}{\partial x^{\nu}}=-\frac{4\pi}{c}j^{\mu}.
\label{CovariantMaxwellOne}%
\end{equation}
[The real part of the first equation traditionally represents the first
(homogeneous) pair of Maxwell's equations and the real part of the second one
gives the second pair. In our approach both pairs of Maxwell's equations
appear together; see also \cite{Barut80}, \cite{Bia:Bia75}, \cite{Bia:Bia13},
\cite{LapUh31}, and \cite{Taylor52} for the case in vacuum. Moreover, a
generalization to complex-valued $4$-current may naturally represent free
magnetic charge and magnetic current not yet observed in nature
\cite{Schwinger}.]

Another important property is a cofactor matrix identity,%
\begin{equation}
P_{\mu\nu}Q^{\nu\lambda}=\left(  \mathbf{F}\cdot\mathbf{G}\right)  \delta
_{\mu}^{\lambda}=\frac{1}{4}\left(  P_{\mu\nu}Q^{\nu\mu}\right)  \delta_{\mu
}^{\lambda}, \label{CofactorIdentity}%
\end{equation}
which was originally established, in a general form, by Minkowski
\cite{MinkowskiI}. Once again, the dual tensors are given by%
\begin{equation}
P_{\mu\nu}=\left(
\begin{array}
[c]{cc}%
0 & F_{q}\medskip\\
-F_{p} & ie_{pqr}G_{r}%
\end{array}
\right)  ,\quad Q^{\mu\nu}=\left(
\begin{array}
[c]{cc}%
0 & -G_{q}\medskip\\
G_{p} & ie_{pqr}F_{r}%
\end{array}
\right)  , \label{DualCompact}%
\end{equation}
in block form. The covariant field energy-momentum tensor in a medium and the
corresponding differential balance equation,%
\begin{align}
&  \frac{\partial}{\partial x^{\nu}}\left[  \frac{1}{16\pi}\left(
P_{\mu\lambda}^{\ast}Q^{\lambda\nu}+P_{\mu\lambda}\overset{\ast}{\left.
Q^{\lambda\nu}\right.  }\right)  \right] \label{EnergyMomentumBalance}\\
&  \ +\frac{1}{32\pi}\left(  P_{\sigma\tau}^{\ast}\frac{\partial Q^{\tau
\sigma}}{\partial x^{\mu}}+P_{\sigma\tau}\frac{\partial\overset{\ast}{\left.
Q^{\tau\sigma}\right.  }}{\partial x^{\mu}}\right)  =\left(
\begin{array}
[c]{c}%
\mathbf{j}\cdot\mathbf{E}/c\smallskip\\
-\rho\mathbf{E}-\mathbf{j}\times\mathbf{B}/c
\end{array}
\right)  ,\nonumber
\end{align}
are derived in terms of these tensors in \cite{KrLanSu15}. (In the rest of the
article, we will be dealing with the case of vacuum only, when $\mathbf{G}%
=\mathbf{F},$ but it's convenient to use both vectors in our calculations
anyway in order to emphasize where they are coming from.)

\section{Transformation Laws and Generators}

Under the Lorentz transformation \cite{Bogolubovetal90}, \cite{MinkowskiI},
\cite{MisThoWhe}, \cite{RumerFetQFT},%
\begin{equation}
U\left(  \Lambda\right)  Q^{\mu\nu}\left(  x^{\rho}\right)  :=Q^{\mu\nu
}\left(  \Lambda_{\rho}^{\kappa}x^{\rho}\right)  =\Lambda_{\sigma}^{\mu
}\Lambda_{\tau}^{\nu}Q^{\sigma\tau}\left(  x^{\rho}\right)
,\label{4TensorTransform}%
\end{equation}
where the summation is assumed over any two repeated indices\footnote{Although
Minkowski considered the transformation of electric and magnetic fields in a
complex $3D$ vector form, see Eqs.~(8)--(9) and (15) in \cite{MinkowskiI} (or
Eqs.~(25.5)--(25.6) in \cite{LanLif2}), he seems never to have combined the
corresponding $4$-tensors into the complex forms~(\ref{ComplexQ}%
)--(\ref{ComplexP}). In the second article \cite{MinkowskiII}, Max Born, who
used Minkowski's notes, didn't mention the complex fields. As a result, the
complex field tensor had appeared, for the first time, in \cite{LapUh31}; see
also \cite{Taylor52}.}. We shall use the following six $4\times4$ matrices
$(\alpha,\beta=0,1,2,3$ are fixed):%
\begin{align}
\Lambda\left(  \theta_{\alpha\beta}\right)   &  =\exp\left(  -\theta
_{\alpha\beta}\ m^{\alpha\beta}\right)  ,\qquad m^{\alpha\beta}=-m^{\beta
\alpha},\label{OneParameter}\\
\left(  m^{\alpha\beta}\right)  _{\nu}^{\mu} &  =g^{\alpha\mu}\delta_{\nu
}^{\beta}-g^{\beta\mu}\delta_{\nu}^{\alpha}\nonumber
\end{align}
for the corresponding one-parameter subgroups of the proper Lorentz group
\cite{Bogolubovetal90}, \cite{MisThoWhe} with the standard metric, $g_{\mu\nu
}=g^{\mu\nu}=$diag$\left(  1,-1,-1,-1\right)  ,$ in the Minkowski space-time.
The $4$-angular momentum operators,%
\begin{equation}
M^{\alpha\beta}=x^{\beta}\partial^{\alpha}-x^{\alpha}\partial^{\beta}%
,\qquad\partial^{\alpha}=g^{\alpha\kappa}\partial_{\kappa}%
,\label{DiffGenerators}%
\end{equation}
can be derived as follows%
\begin{align}
M^{\alpha\beta}Q^{\mu\nu} &  :=-\left.  \left[  \frac{d}{d\theta_{\alpha\beta
}}Q^{\mu\nu}\left(  \Lambda_{\rho}^{\kappa}\left(  \theta_{\alpha\beta
}\right)  x^{\rho}\right)  \right]  \right\vert _{\theta_{\alpha\beta}%
=0}\label{Generators}\\
&  =\left(  m^{\alpha\beta}\right)  _{\sigma}^{\mu}Q^{\sigma\nu}+Q^{\mu\tau
}\left(  m^{\alpha\beta}\right)  _{\tau}^{\nu}\nonumber
\end{align}
with%
\begin{equation}
\left(  m^{\alpha\beta}\right)  _{\nu}^{\mu}=-\left.  \frac{d\Lambda_{\nu
}^{\mu}\left(  \theta_{\alpha\beta}\right)  }{d\theta_{\alpha\beta}%
}\right\vert _{\theta_{\alpha\beta}=0}=g^{\alpha\mu}\delta_{\nu}^{\beta
}-g^{\beta\mu}\delta_{\nu}^{\alpha}.\label{GeneratorMatrices}%
\end{equation}
In matrix form,%
\begin{equation}
M^{\alpha\beta}Q=m^{\alpha\beta}Q+Q\left(  m^{\alpha\beta}\right)
^{T},\label{GeneratorsMatrixForm}%
\end{equation}
where $Q=Q^{\mu\nu}$ and $m^{T}$ is the transposed matrix. The latter
equations (\ref{Generators}), (\ref{GeneratorMatrices}), and
(\ref{GeneratorsMatrixForm}) define the action of the infinitesimal operators
on the complex field tensor,%
\begin{equation}
M^{\sigma\tau}Q^{\alpha\beta}=g^{\sigma\alpha}Q^{\tau\beta}-g^{\tau\alpha
}Q^{\sigma\beta}+g^{\sigma\beta}Q^{\alpha\tau}-g^{\tau\beta}Q^{\alpha\sigma
},\label{GeneratorAction}%
\end{equation}
in the form that is required in equation (\ref{PauliLubanski}) below.

In a similar fashion, for the products of the generators,%
\begin{align}
M^{\alpha\beta}M^{\gamma\delta}Q^{\mu\nu}  &  =\left(  m^{\gamma\delta
}\right)  _{\kappa}^{\mu}\left(  m^{\alpha\beta}\right)  _{\sigma}^{\kappa
}Q^{\sigma\nu}+\left(  m^{\alpha\beta}\right)  _{\sigma}^{\mu}\left(
m^{\gamma\delta}\right)  _{\rho}^{\nu}Q^{\sigma\rho} \label{GeneratorProducts}%
\\
&  +\left(  m^{\gamma\delta}\right)  _{\kappa}^{\mu}\left(  m^{\alpha\beta
}\right)  _{\tau}^{\nu}Q^{\kappa\tau}+\left(  m^{\gamma\delta}\right)  _{\rho
}^{\nu}\left(  m^{\alpha\beta}\right)  _{\tau}^{\rho}Q^{\mu\tau},\nonumber
\end{align}
or, in matrix form,%
\begin{align}
M^{\alpha\beta}M^{\gamma\delta}Q  &  =\left(  m^{\gamma\delta}m^{\alpha\beta
}\right)  Q-\left(  \left(  m^{\gamma\delta}m^{\alpha\beta}\right)  Q\right)
^{T}\label{GeneratorProductMatrix}\\
&  +m^{\alpha\beta}Q\left(  m^{\gamma\delta}\right)  ^{T}-\left(
m^{\alpha\beta}Q\left(  m^{\gamma\delta}\right)  ^{T}\right)  ^{T}.\nonumber
\end{align}
As a result,%
\begin{align}
&  \left[  M^{\alpha\beta},\ M^{\gamma\delta}\right]  :=M^{\alpha\beta
}M^{\gamma\delta}-M^{\gamma\delta}M^{\alpha\beta}\label{Commutators}\\
&  \quad=g^{\alpha\gamma}M^{\beta\delta}-g^{\alpha\delta}M^{\beta\gamma
}+g^{\beta\delta}M^{\alpha\gamma}-g^{\beta\gamma}M^{\alpha\delta},\nonumber
\end{align}
which follows from (\ref{Generators})--(\ref{GeneratorMatrices}) and can be
verified, once again, by using (\ref{DiffGenerators}).

Finally, introducing the infinitesimal operators $\mathbf{M}=\left(
M^{23},M^{31},M^{12}\right)  $ and $\mathbf{N}=\left(  M^{01},M^{02}%
,M^{03}\right)  $ for the rotations and boosts, respectively, one can get%
\begin{equation}
\mathbf{N}^{2}Q=-\mathbf{M}^{2}Q=2Q,\qquad\left(  \mathbf{M}\cdot
\mathbf{N}\right)  Q=-2iQ. \label{CasimirLorentz}%
\end{equation}
The Casimir operators of the proper Lorentz group are given by $\left(
\mathbf{M}+i\mathbf{N}\right)  ^{2}/4=0$ and $\left(  \mathbf{M-}%
i\mathbf{N}\right)  ^{2}/4=-2$ in the space of complex anti-symmetric tensors
under consideration. In view of $\mathbf{M}^{2}=-s\left(  s+1\right)  =-2,$ we
may say that the spin of the photon is equal to one. (Here, we have chosen
anti-hermitian operators; see also \cite{BargmannWigner48}, \cite{Gelfandetal}%
, \cite{RumerFetQFT}, and \cite{Wein} for more details on the Lorentz group representations.)

\section{The Pauli-Luba\'{n}ski Vector and Maxwell's Equations in Vacuum}

As follows from the representation theory of the Poincar\'{e} group
\cite{BargmannWigner48}, \cite{Bogolubovetal90} and the geometry of the
Minkowski space-time \cite{Minkowski}, \cite{Naber12}, for the case of
massless particles, the Pauli-Luba\'{n}ski vector should be collinear to the
operator of the $4$-linear momentum. For a classical electromagnetic field,
this relation takes the form%
\begin{equation}
\frac{1}{2}e_{\mu\nu\sigma\tau}\partial^{\nu}\left(  M^{\sigma\tau}%
Q^{\alpha\beta}\right)  =-i\partial_{\mu}Q^{\alpha\beta},
\label{PauliLubanski}%
\end{equation}
and by (\ref{GeneratorAction}), we find that%
\begin{equation}
g^{\alpha\alpha}e_{\alpha\mu\nu\tau}\partial^{\nu}Q^{\tau\beta}-g^{\beta\beta
}e_{\beta\mu\nu\tau}\partial^{\nu}Q^{\tau\alpha}=-i\partial_{\mu}%
Q^{\alpha\beta} \label{KLS}%
\end{equation}
($\alpha,\beta=0,1,2,3$ are fixed; no summation is assumed over these two
indices). By a direct, but rather tedious evaluation, one can verify that the
latter equation, which is written in terms of a third rank $4$-tensor, is
equivalent to the original system of Maxwell equations in vacuum,
$\partial_{\nu}Q^{\mu\nu}=0.$ As a result, the helicity of the
photon\footnote{Multiple meanings of the word \textquotedblleft
photon\textquotedblright\ are analized in \cite{Klyshko94}.}, or a harmonic
circular classical electromagnetic wave, cannot be defined as an undetermined
sign, or an extra $\pm1$ factor, in the right hand side of equation
(\ref{PauliLubanski}) as it is stated in standard textbooks on the quantum
field theory \cite{BargmannWigner48}, \cite{Bogolubovetal90}, \cite{Ryder},
\cite{ScheckQuantumPhysics}. (This misconception has been one of our main
motivations for writing this article.)

In view of (\ref{PauliLubanski}), for the rotations and boosts, $\mathbf{M}%
=\left(  M^{23},M^{31},M^{12}\right)  $ and $\mathbf{N}=\left(  M^{01}%
,M^{02},M^{03}\right)  ,$ respectively, the following standard equations hold%
\begin{equation}
\left(  \nabla\cdot\mathbf{M}\right)  Q=i\partial_{0}Q,\quad\partial_{0}%
=\frac{1}{c}\frac{\partial}{\partial t} \label{KLSzero}%
\end{equation}
and%
\begin{equation}
\partial_{0}\mathbf{M}Q+\left(  \nabla\times\mathbf{N}\right)  Q=i\nabla Q,
\label{KLSvector}%
\end{equation}
where $Q=Q^{\alpha\beta}=-Q^{\beta\alpha}$ is the complex field tensor and the
actions of operators $\mathbf{M}$ and $\mathbf{N}$ on this tensor are
explicitly defined by (\ref{GeneratorAction}).

\noindent\textbf{Note.} In vacuum, when $\mathbf{G}=\mathbf{F}$ and $\rho=0,$
$\mathbf{j}=0,$ two different covariant forms of Maxwell's equations are given
by%
\begin{equation}
\partial_{\nu}Q^{\mu\nu}=0,\qquad\partial^{\nu}P_{\mu\nu}=0,
\label{MaxwellPQinVacuum}%
\end{equation}
where $\partial^{\nu}=g^{\nu\mu}\partial_{\mu}=g^{\nu\mu}\partial/\partial
x^{\mu}.$ The second equation follows from (\ref{KLS}), when one takes
$\mu=\beta$ and sums over $\beta=0,1,2,3$ with the help of (\ref{DualPQ}). As
another useful consequence of our equation (\ref{KLS}), one can directly show
that the d'Alembert operator annihilates any component of the complex field
tensor in vacuum,%
\begin{equation}
\partial^{\mu}\partial_{\mu}Q^{\alpha\beta}=\left(  \frac{1}{c^{2}}%
\frac{\partial^{2}}{\partial t^{2}}-\Delta\right)  Q^{\alpha\beta}=\square
Q^{\alpha\beta}=0, \label{dAlembert}%
\end{equation}
thus de-coupling the system. It is worth noting that, in covariant form, our
derivation does not require any formula of $3D$-vector calculus. (The general
theory of relativistic-invariant equations is studied in
\cite{BargmannWigner48}, \cite{Gelfandetal}; see also \cite{Bia:Bia75},
\cite{GolSht01}, \cite{Lub41}, \cite{Lub42-I}, \cite{Lub42-II} and the
references therein.)

\section{Examples}

In a matrix form, equation (\ref{KLSzero}) can be rewritten as follows%
\begin{align}
&  \left(
\begin{array}
[c]{cccc}%
0 & -\partial_{2}G_{3}+\partial_{2}G_{3} & \partial_{1}G_{3}-\partial_{3}G_{1}
& -\partial_{1}G_{2}+\partial_{2}G_{1}\smallskip\\
\partial_{2}G_{3}-\partial_{2}G_{3} & 0 & i\left(  \partial_{1}F_{2}%
-\partial_{2}F_{1}\right)  & i\left(  \partial_{1}F_{3}-\partial_{3}%
F_{1}\right)  \smallskip\\
-\partial_{1}G_{3}+\partial_{3}G_{1} & -i\left(  \partial_{1}F_{2}%
-\partial_{2}F_{1}\right)  & 0 & i\left(  \partial_{2}F_{3}-\partial_{3}%
F_{2}\right)  \smallskip\\
\partial_{1}G_{2}-\partial_{2}G_{1}\smallskip & -i\left(  \partial_{1}%
F_{3}-\partial_{3}F_{1}\right)  & -i\left(  \partial_{2}F_{3}-\partial
_{3}F_{2}\right)  & 0
\end{array}
\right) \nonumber\\
&  \quad=+\frac{i}{c}\frac{\partial}{\partial t}\left(
\begin{array}
[c]{cccc}%
0 & -G_{1} & -G_{2} & -G_{3}\smallskip\\
G_{1} & 0 & iF_{3} & -iF_{2}\smallskip\\
G_{2} & -iF_{3} & 0 & iF_{1}\smallskip\\
G_{3} & iF_{2}\smallskip & -iF_{1} & 0
\end{array}
\right)  , \label{Example0}%
\end{align}
or%
\begin{equation}
\left(
\begin{array}
[c]{cc}%
0 & -\left(  \operatorname{curl}\mathbf{G}\right)  _{q}\medskip\\
\left(  \operatorname{curl}\mathbf{G}\right)  _{p} & ie_{pqr}\left(
\operatorname{curl}\mathbf{F}\right)  _{r}%
\end{array}
\right)  =\frac{i}{c}\frac{\partial}{\partial t}\left(
\begin{array}
[c]{cc}%
0 & -G_{q}\medskip\\
G_{p} & ie_{pqr}F_{r}%
\end{array}
\right)  , \label{Example0Compact}%
\end{equation}
in a more compact block form. In vacuum, $\mathbf{G}=\mathbf{F}$ and this
matrix relation implies the single complex Maxwell equation,
$\operatorname{curl}\mathbf{F}=\left(  i/c\right)  \partial\mathbf{F}/\partial
t.$

In a similar fashion, for the first component of (\ref{KLSvector}), namely,
$\partial_{0}M_{1}Q+\left(  \partial_{2}N_{3}-\partial_{3}N_{2}\right)
Q=i\partial_{1}Q,$ we obtain,%
\begin{align}
&  \partial_{0}\left(
\begin{array}
[c]{cccc}%
0 & 0 & G_{3} & -G_{2}\smallskip\\
0 & 0 & iF_{2} & iF_{3}\smallskip\\
-G_{3} & -iF_{2} & 0 & 0\smallskip\\
G_{2} & -iF_{3} & 0 & 0
\end{array}
\right) \label{Example1}\\
&  +\left(
\begin{array}
[c]{cccc}%
0 & i\left(  \partial_{2}F_{2}+\partial_{3}F_{3}\right)  & -i\partial_{2}F_{1}
& -i\partial_{3}F_{1}\smallskip\\
-i\left(  \partial_{2}F_{2}+\partial_{3}F_{3}\right)  & 0 & -\partial_{3}G_{1}
& \partial_{2}G_{1}\smallskip\\
i\partial_{2}F_{1} & \partial_{3}G_{1} & 0 & \partial_{2}G_{2}+\partial
_{3}G_{3}\smallskip\\
i\partial_{3}F_{1}\smallskip & -\partial_{2}G_{1}\smallskip & -\partial
_{2}G_{2}-\partial_{3}G_{3} & 0
\end{array}
\right) \nonumber\\
&  \quad=+i\partial_{1}\left(
\begin{array}
[c]{cccc}%
0 & -G_{1} & -G_{2} & -G_{3}\smallskip\\
G_{1} & 0 & iF_{3} & -iF_{2}\smallskip\\
G_{2} & -iF_{3} & 0 & iF_{1}\smallskip\\
G_{3} & iF_{2}\smallskip & -iF_{1}\smallskip & 0
\end{array}
\right)  .\nonumber
\end{align}
Once again, in vacuum, $\mathbf{G}=\mathbf{F}$ and this matrix relation is
satisfied in view of the pair of complex Maxwell equations,
$\operatorname{curl}\mathbf{F}=\left(  i/c\right)  \partial\mathbf{F}/\partial
t$ and $\operatorname{div}\mathbf{F}=0.$ (A cyclic permutation of the spatial
indices covers the two remaining components.)

One can clearly see that there is no chance of changing the sign $+$ into $-$
in the right hand side without a violation of Maxwell's equations. Indeed, let
us pick just one of the matrix elements from both sides, say, $\partial
_{2}F_{2}+\partial_{3}F_{3}=-\partial_{1}G_{1},$ which indicates also that the
left and right hand sides are coming from the different pairs of Maxwell's
equations (\ref{MaxwellOne})--(\ref{MaxwellTwo}).

\section{Helicity}

In particle physics \cite{Ber:Lif:Pit}, \cite{RumerFetQFT}, \cite{TroTy94},
the helicity is defined as the projection of the angular momentum $\mathbf{M}$
on the direction of motion $\mathbf{p}:$
\begin{equation}
\lambda=\frac{\mathbf{p}\cdot\mathbf{M}}{\left\vert \mathbf{p}\right\vert
}=-\frac{w_{0}}{\left\vert \mathbf{p}\right\vert }. \label{HelicityOperator}%
\end{equation}
The helicity states are eigenstates of the operator:%
\begin{equation}
\Lambda\left\vert \mathbf{p},\lambda\right\rangle =\frac{\mathbf{p}%
\cdot\mathbf{M}}{\left\vert \mathbf{p}\right\vert }\left\vert \mathbf{p}%
,\lambda\right\rangle =\lambda\left\vert \mathbf{p},\lambda\right\rangle .
\label{HelicityEigenstates}%
\end{equation}
For massless particles one can define the spin as $s=\left\vert \lambda
\right\vert $ and, if the parity is conserved, the particle may have only two
independent helicity eigenstates $\left\vert \mathbf{p},\lambda=s\right\rangle
$ and $\left\vert \mathbf{p},\lambda=-s\right\rangle .$

In the case of the classical electromagnetic field, equations (\ref{KLSzero})
and (\ref{Example0Compact}) together show that the helicity operator is
proportional to the \textquotedblleft energy operator\textquotedblright:%
\begin{equation}
\Lambda=\frac{i}{c\left\vert \mathbf{k}\right\vert }\frac{\partial}{\partial
t}. \label{HelicityEnergyOperator}%
\end{equation}
As a result, these two operators have common eigenstates, $\left\vert
\mathbf{k},\lambda\right\rangle =Q^{\mu\nu},$ in the space of complex
anti-symmetric $4$-tensors of the second rank. (The simplest covariant
helicity states will be constructed in the next section.)

On the other hand, in $3D$-complex electrodynamics, one can take the complex
vector field $\left\vert \mathbf{k},\lambda\right\rangle =\mathbf{F}%
=\mathbf{E}+i\mathbf{H}$ and choose the following real-valued spin matrices
\cite{VMK}:%
\begin{equation}
s_{1}=\left(
\begin{array}
[c]{ccc}%
0 & 0 & 0\\
0 & 0 & -1\\
0 & 1 & 0
\end{array}
\right)  ,\quad s_{2}=\left(
\begin{array}
[c]{ccc}%
0 & 0 & 1\\
0 & 0 & 0\\
-1 & 0 & 0
\end{array}
\right)  ,\quad s_{3}=\left(
\begin{array}
[c]{ccc}%
0 & -1 & 0\\
1 & 0 & 0\\
0 & 0 & 0
\end{array}
\right)  ,\quad\label{SpinMatrices}%
\end{equation}
such that $\left[  s_{p},s_{q}\right]  =s_{p}s_{q}-s_{q}s_{p}=e_{pqr}s_{r}$
and $s_{1}^{2}+s_{2}^{2}+s_{3}^{2}=-2.$ Then%
\begin{align}
&  \left(  \nabla\cdot\mathbf{s}\right)  \mathbf{F}:=\partial_{1}\left(
s_{1}\mathbf{F}\right)  +\partial_{2}\left(  s_{2}\mathbf{F}\right)
+\partial_{3}\left(  s_{3}\mathbf{F}\right) \label{SpinCurl}\\
&  \ =\partial_{1}\left(
\begin{array}
[c]{ccc}%
0 & 0 & 0\\
0 & 0 & -1\\
0 & 1 & 0
\end{array}
\right)  \left(
\begin{array}
[c]{c}%
F_{1}\\
F_{2}\\
F_{3}%
\end{array}
\right)  +\partial_{2}\left(
\begin{array}
[c]{ccc}%
0 & 0 & 1\\
0 & 0 & 0\\
-1 & 0 & 0
\end{array}
\right)  \left(
\begin{array}
[c]{c}%
F_{1}\\
F_{2}\\
F_{3}%
\end{array}
\right) \nonumber\\
&  \ \ \ +\partial_{3}\left(
\begin{array}
[c]{ccc}%
0 & -1 & 0\\
1 & 0 & 0\\
0 & 0 & 0
\end{array}
\right)  \left(
\begin{array}
[c]{c}%
F_{1}\\
F_{2}\\
F_{3}%
\end{array}
\right)  =\left(
\begin{array}
[c]{c}%
\partial_{2}F_{3}-\partial_{3}F_{2}\\
\partial_{3}F_{1}-\partial_{1}F_{3}\\
\partial_{1}F_{2}-\partial_{2}F_{1}%
\end{array}
\right)  =\operatorname{curl}\mathbf{F}.\nonumber
\end{align}
Once again, our representation (\ref{HelicityEnergyOperator}) for the helicity
operator holds in view of the Maxwell equation in vacuum, $\operatorname{curl}%
\mathbf{F}=\left(  i/c\right)  \partial\mathbf{F}/\partial t.$

\noindent\textbf{Note.} In view of (\ref{HelicityEnergyOperator}), the
traditional definition of helicity (\ref{HelicityEigenstates}) is related to
separation of variables in Maxwell's equations. Letting $Q^{\mu\nu}=q\left(
t\right)  Z^{\mu\nu}\left(  \mathbf{r}\right)  ,$ one gets%
\begin{equation}
0=\frac{\partial Q^{\mu\nu}}{\partial x^{\nu}}=\frac{1}{c}\frac{\partial
Q^{\mu0}}{\partial t}+\frac{\partial Q^{\mu p}}{\partial x_{p}}=\frac{1}%
{c}\overset{\mathbf{\cdot}}{q}\left(  t\right)  Z^{\mu0}\left(  \mathbf{r}%
\right)  +q\left(  t\right)  \frac{\partial Z^{\mu p}\left(  \mathbf{r}%
\right)  }{\partial x_{p}},
\end{equation}
or%
\begin{equation}
-\frac{\overset{\mathbf{\cdot}}{q}}{cq}Z^{\mu0}=\frac{\partial Z^{\mu p}%
}{\partial x_{p}},\qquad q=e^{-i\omega t},
\end{equation}
where $\omega$ must be a real-valued constant of the separation of variables
in order to have bounded solutions. As a result,%
\begin{equation}
\frac{\partial Z^{\mu p}}{\partial x_{p}}=i\frac{\omega}{c}Z^{\mu0},
\end{equation}
thus giving a covariant form of the corresponding eigenvalue problems in
different curvilinear coordinates \cite{ToptyginI}, \cite{VMK}.

\section{Covariant Harmonic Wave Solutions}

In vacuum, $\partial_{\nu}Q^{\mu\nu}=0,$ where%
\begin{equation}
Q^{\mu\nu}=\left(
\begin{array}
[c]{cc}%
0 & -F_{q}\medskip\\
F_{p} & ie_{pqr}F_{r}%
\end{array}
\right)  ,\quad\mathbf{F}=\mathbf{f}e^{i\left(  \omega t-\mathbf{k}%
\cdot\mathbf{r}\right)  }=\mathbf{E}+i\mathbf{H}. \label{HarmonicWave}%
\end{equation}
Here, $\mathbf{f}=$constant is a complex polarization vector to be determined
and%
\begin{equation}
x^{\mu}=\left(  ct,\mathbf{r}\right)  ,\qquad k_{\mu}=\left(  \omega
/c,-\mathbf{k}\right)  ,\qquad kx=k_{\mu}x^{\mu}=\omega t-\mathbf{k}%
\cdot\mathbf{r}. \label{WaveVector}%
\end{equation}
In a compact form, $Q^{\mu\nu}=A^{\mu\nu}e^{ikx}$ and $A^{\mu\nu}k_{\nu
}=0^{\mu},$ where
\begin{equation}
A^{\mu\nu}=\left(
\begin{array}
[c]{cc}%
0 & -f_{q}\medskip\\
f_{p} & ie_{pqr}f_{r}%
\end{array}
\right)  =\text{constant}. \label{PolarizationTensor}%
\end{equation}
This tensor is an eigenfunction of the $4$-gradient, $i^{-1}\partial_{\alpha
}Q^{\mu\nu}=k_{\alpha}Q^{\mu\nu}.$

As a result,%
\begin{equation}
\left(
\begin{array}
[c]{cccc}%
0 & -f_{1} & -f_{2} & -f_{3}\\
f_{1} & 0 & if_{3} & -if_{2}\\
f_{2} & -if_{3} & 0 & if_{1}\\
f_{3} & if_{2} & -if_{1} & 0
\end{array}
\right)  \left(
\begin{array}
[c]{c}%
\omega/c\\
-k_{1}\\
-k_{2}\\
-k_{3}%
\end{array}
\right)  =\left(
\begin{array}
[c]{c}%
0\\
0\\
0\\
0
\end{array}
\right)  \label{KernelMatrix}%
\end{equation}
and $\det A=-\left(  \mathbf{f}\cdot\mathbf{f}\right)  ^{2}=0$ (Lorentz
invariant by Minkowski \cite{MinkowskiI}). The complex invariant,
$\mathbf{F}^{2}=\left(  \mathbf{E}+i\mathbf{H}\right)  ^{2}=0,$ results in
$\mathbf{E}^{2}=\mathbf{H}^{2}$ and $\mathbf{E}\cdot\mathbf{H}=0,$ as required
\cite{LanLif2}.

In $3D$-form, the latter system of linear equations gives an eigenvalue
problem:%
\begin{equation}
i\mathbf{k}\times\mathbf{f}=\frac{\omega}{c}\mathbf{f},\qquad\mathbf{f}%
\cdot\mathbf{f}=0. \label{EigenvalueProblem}%
\end{equation}
The eigenvalues are%
\begin{equation}
\left\vert
\begin{array}
[c]{ccc}%
-\omega/c & -ik_{3} & ik_{2}\\
ik_{3} & -\omega/c & -ik_{1}\\
-ik_{2} & ik_{1} & -\omega/c
\end{array}
\right\vert =\frac{\omega}{c}\left(  k_{1}^{2}+k_{2}^{2}+k_{3}^{2}%
-\frac{\omega^{2}}{c^{2}}\right)  =0. \label{Eigenvalues}%
\end{equation}
The case $\omega=0,$ when $\mathbf{f}=\mathbf{k},$ does not satisfy the second
condition $\mathbf{f}^{2}=0$ unless $\mathbf{k}=\mathbf{0}.$

Therefore, there are only two eigenvectors $\left\{  \mathbf{f},\ \mathbf{f}%
^{\ast}\right\}  $, corresponding to $\omega/c=\pm k=\pm\sqrt{k_{1}^{2}%
+k_{2}^{2}+k_{3}^{2}}:$%
\begin{equation}
\mathbf{f}=\frac{\mathbf{k}\times\left(  \boldsymbol{l}\times\mathbf{k}%
\right)  +ik\left(  \mathbf{k}\times\boldsymbol{l}\right)  }{k\sqrt
{2}\left\vert \mathbf{k}\times\boldsymbol{l}\right\vert },\qquad
\mathbf{f}^{\ast}=\left.  \mathbf{f}\right\vert _{\mathbf{k}\rightarrow
-\mathbf{k}}, \label{PolarizationVectors}%
\end{equation}
respectively \cite{Bia:Bia13}. Here, $\boldsymbol{l}$ is an arbitrary real
vector that is not collinear to $\mathbf{k}$ ($\mathbf{k}\neq$%
constant$\ \boldsymbol{l}$) and $\mathbf{f}\cdot\mathbf{f}^{\ast}=1.$ (A
similar eigenvalue problem occurs in the mean magnetic field generation,
called $\alpha\Omega$-dynamo, in cosmic astrophysics \cite{FleishTop13}.)

\noindent{\textbf{Example}}. Let $\left\{  \mathbf{e}_{k}\right\}  _{k=1}^{3}$
be an orthonormal basis in $\left.
\mathbb{R}
\right.  ^{3}.$ One can choose $\boldsymbol{l}=\mathbf{e}_{1}$ and
$\mathbf{k}=k\mathbf{e}_{3}.$ Then%
\begin{equation}
\mathbf{f}=\frac{\mathbf{e}_{1}+i\mathbf{e}_{2}}{\sqrt{2}},\qquad
\mathbf{f}^{\ast}=\frac{\mathbf{e}_{1}-i\mathbf{e}_{2}}{\sqrt{2}}
\label{PolarizationVectorsExample}%
\end{equation}
(see \cite{Bia:Bia13}, \cite{LanLif2}, and \cite{ToptyginI} for more details).

\section{Discrete Transformations and Polarization}

The complex Maxwell equations in vacuum,%
\begin{equation}
\frac{i}{c}\frac{\partial\mathbf{F}}{\partial t}=\operatorname{curl}%
\mathbf{F},\qquad\operatorname{div}\mathbf{F}=0, \label{ComplexMaxwell}%
\end{equation}
are invariant under the following discrete transformations: spatial inversion
\textbf{P:\ }$\mathbf{F}\left(  \mathbf{r},t\right)  \rightarrow
\mathbf{F}^{\ast}\left(  -\mathbf{r},t\right)  ;$ time reversal \textbf{T}%
$:\ \mathbf{F}\left(  \mathbf{r},t\right)  \rightarrow\mathbf{F}^{\ast}\left(
\mathbf{r},-t\right)  ;$ and space-time inversion \textbf{PT}$:\ \mathbf{F}%
\left(  \mathbf{r},t\right)  \rightarrow\mathbf{F}\left(  -\mathbf{r}%
,-t\right)  .$ They, together with the identity transformation, correspond to
the four connected components of the Poincar\'{e} group. (These
transformations form the so-called Klein group $\left\{  \mathbf{Identity}%
,\mathbf{P},\mathbf{T},\mathbf{PT}\right\}  .$)

The action of this group generates the following four circularly polarized
waves ($\omega=+ck$):%
\begin{equation}
\mathbf{F}_{1}=\mathbf{f}e^{i\left(  \mathbf{k}\cdot\mathbf{r}-\omega
t\right)  },\quad\mathbf{F}_{2}=\mathbf{f}^{\ast}e^{-i\left(  \mathbf{k}%
\cdot\mathbf{r}+\omega t\right)  }=\left.  \mathbf{F}_{1}^{\ast}\right\vert
_{t\rightarrow-t}=\mathbf{TF}_{1} \label{PositiveHelicity}%
\end{equation}
and%
\begin{align}
\mathbf{F}_{3}  &  =\mathbf{f}e^{-i\left(  \mathbf{k}\cdot\mathbf{r}-\omega
t\right)  }=\left.  \mathbf{F}_{1}\right\vert _{\mathbf{r}\rightarrow
-\mathbf{r},t\rightarrow-t}=\left(  \mathbf{PT}\right)  \mathbf{F}%
_{1},\label{NegativeHelicity}\\
\mathbf{F}_{4}  &  =\mathbf{f}^{\ast}e^{i\left(  \mathbf{k}\cdot
\mathbf{r}+\omega t\right)  }=\left.  \mathbf{F}_{3}^{\ast}\right\vert
_{t\rightarrow-t}=\mathbf{TF}_{3}=\mathbf{PF}_{1}.\nonumber
\end{align}
They represent right- and left-handed circularly polarized waves moving along
the vector $\mathbf{k}$ in opposite directions. One can easily verify that the
solutions $\left\{  \mathbf{F}_{1},\mathbf{F}_{2}\right\}  $ correspond to
$\lambda=+1$ and $\left\{  \mathbf{F}_{3},\mathbf{F}_{4}\right\}  $ have
$\lambda=-1.$ Also, $\mathbf{F}_{1}\cdot\mathbf{F}_{3}=\mathbf{f}^{2}=0$ and
$\mathbf{F}_{2}\cdot\mathbf{F}_{4}=\left(  \mathbf{f}^{2}\right)  ^{\ast}=0.$

\noindent{\textbf{Example}}. The standard circular, elliptic, and linear
polarizations of the classical electromagnetic waves occur as a result of
superposition of the complex solutions under consideration. With the help of
the polarization vectors (\ref{PolarizationVectorsExample}), one gets%
\begin{align}
&  \mathbf{F}=c_{1}\frac{\mathbf{F}_{1}+\mathbf{F}_{3}}{2}+c_{2}%
\frac{\mathbf{F}_{1}-\mathbf{F}_{3}}{2}=\mathbf{E}+i\mathbf{H}%
\label{EllipticPolarization}\\
&  \ \ =c_{1}\mathbf{e}_{1}\cos\left(  \mathbf{k}\cdot\mathbf{r}-\omega
t\right)  -c_{2}\mathbf{e}_{2}\sin\left(  \mathbf{k}\cdot\mathbf{r}-\omega
t\right) \nonumber\\
&  \ \ +i\left[  c_{2}\mathbf{e}_{1}\cos\left(  \mathbf{k}\cdot\mathbf{r}%
-\omega t-\frac{\pi}{2}\right)  -c_{1}\mathbf{e}_{2}\sin\left(  \mathbf{k}%
\cdot\mathbf{r}-\omega t-\frac{\pi}{2}\right)  \right]  ,\nonumber
\end{align}
where $c_{1}=c_{1}^{\ast}$ and $c_{2}=c_{2}^{\ast}.$ For the elliptic
polarization, we choose $\left\vert c_{1}\right\vert >\left\vert
c_{2}\right\vert $ or $\left\vert c_{1}\right\vert <\left\vert c_{2}%
\right\vert ;$ the linear polarization arises, for instance, if $c_{1}\neq0$
and $c_{2}=0$ (see \cite{LanLif2} and \cite{ToptyginI}, problems~2.128--2.134,
for more details).

In conclusion, it is worth noting that, here, we have only discussed the
classical electromagnetic field in vacuum. Different aspects of the
\textquotedblleft photon paradigm\textquotedblright\ are emphasized in
\cite{Klyshko94}. The photon wave functions are dealt with in \cite{Akh:Ber},
\cite{Ber:Lif:Pit}, \cite{Bia94}, \cite{Bia96}, \cite{DutraQED},
\cite{Fock30-5}. For quantization in the complex form, see \cite{Bia94},
\cite{Bia06}, \cite{Bia:Bia75}, \cite{Bia:Bia13} and the references therein.
(General quantization procedures are discussed, for example, in
\cite{Bia:Bia75}, \cite{Bogolubovetal90}, \cite{KlauderSudarshan},
\cite{Kr:Sus12}, \cite{KrSuaSu14}, \cite{MoskalevQFT}, \cite{MukhWin07},
\cite{Ryder}, \cite{ScheckQuantumPhysics}, \cite{ScullyZub}, \cite{ToptyginI}%
.) Coherent states of light and dynamical invariants are reviewed in
\cite{Dod:Mal:Man75}, \cite{Dodonov:Man'koFIAN87}, \cite{DodonovManko03},
\cite{KlauderSudarshan}, \cite{ScullyZub}. The squeezed states of the photons
and atoms in a cavity and their relations with so-called \textquotedblleft
missing\textquotedblright\ solutions for the harmonic oscillator are analyzed
in \cite{KrySusVegaMinimum}, \cite{Lop:Sus:VegaGroup}, \cite{LopSusVegaHarm}.
Professor Toptygin kindly pointed out an intrinsic importance of the helicity
concept on an enormous scale, from the sub-atomic world (parity violation in
beta decay \cite{Roy01}, \cite{Wuetal57}) to cosmic astrophysics (possible
amplification of galactic magnetic fields by the turbulent dynamo mechanism
\cite{Beck09}, \cite{FleishTop13}, \cite{Vallee04}). Last but not least,
organic compounds appear often in the form of only one of two stereoisomers.
As a result, in optically active biological substances, these molecules rotate
polarized light to the left \cite{Hecht02}, thus creating another old
unexplained puzzle.

\noindent{\textbf{Acknowledgements}}. This research was partially supported by
NSF grant DMS~\#~1535822. We are grateful to Prof. Dr. Patric Muggli for his
hospitality at Max-Planck-Institut f\"{u}r Physik, Werner-Heisenberg-Institute
in Munich; March 2015. We are indebted to Prof. Dr. Gerald A. Goldin, Prof.
Dr.~John Klauder, Prof. Dr. Margarita A. Man'ko, Prof. Dr. Vladimir I. Man'ko,
Prof. Dr. Svetlana Roudenko, and Prof. Dr. Igor N. Toptygin for their valuable
comments and encouragement. The third-named author was partially supported by
the AFOSR grant FA9550-11-1-0220. Comments from one of the referees, which
have helped to improve the presentation, are appreciated.

\end{document}